# Deep Learning for Encrypted Traffic Classification and Unknown Data Detection

Madushi H. Pathmaperuma, Yogachandran Rahulamathavan, *Senior Member, IEEE,* Safak Dogan, *Senior Member, IEEE,* Ahmet M. Kondoz, *Senior Member, IEEE,* and Rongxing Lu, *Fellow, IEEE,*

*Abstract—* Despite the widespread use of encryption techniques to provide confidentiality over Internet communications, mobile device users are still susceptible to privacy and security risks. In this paper, a new Deep Neural Network (DNN) based user activity detection framework is proposed to identify fine grained user activities performed on mobile applications (known as in-app activities) from a sniffed encrypted Internet traffic stream. One of the challenges is that there are countless applications, and it is practically impossible to collect and train a DNN model using all possible data from them. Therefore, in this work we exploit the probability distribution of DNN output layer to filter the data from applications that are not considered during the model training (i.e., unknown data). The proposed framework uses a time window based approach to divide the traffic flow of an activity into segments, so that in-app activities can be identified just by observing only a fraction of the activity related traffic. Our tests have shown that the DNN based framework has demonstrated an accuracy of 90% or above in identifying previously trained in-app activities and an average accuracy of 79% in identifying previously untrained in-app activity traffic as unknown data when this framework is employed.

## I. INTRODUCTION

SMARTPHONES have increasingly become the primary user devices for accessing the Internet in recent years. Both mobile websites and applications (aka apps) can be accessed via a mobile device. Usually, users prefer using apps installed on their devices instead of accessing the corresponding websites from web browsers due to better personalization, speed and easy access offered by mobile apps. Thus, web browsers have largely been replaced by mobile apps for interacting with most online services such as social networking, finances, and media streaming. According to [1], it is shown that mobile users spend nearly 90% of their time on mobile apps compared to websites.

Mobile devices are often connected to IEEE 802.11 Wire-less Local Area Networks (WLAN). Wireless networks employ security protocols such as Wired Equivalent Privacy (WEP), Wi-Fi Protected Access (WPA), WPA2 and WPA3 to secure the wireless connection and most networks are protected by strong passwords [2]. Even though these security protocols are used to encrypt data, it is still possible to passively sniff wireless network traffic and infer sensitive information of users such as the activities performed within apps, location of the user, websites visited by the user etc [3]. This is because even though encryption protects the packet's payload, it does not hide

---------------------------------------------------------------

information revealed by network traffic patterns such as frame length, inter arrival time and direction (incoming/ outgoing). These are known as *side channel data* and can be used to reveal private information related to user's online behaviour.

Users perform various activities using apps installed on their mobile devices. Different in-app activities have different network behaviours [3], generating different traffic flow statistics such as maximum frame length, mean interarrival time etc. These statistical features can be used to fingerprint the in-app activities and later identify them in generic network traces.

Network traffic classification methods have evolved significantly over time from port based, Deep Packet Inspection (DPI) to Machine Learning (ML) methods. Increasing use of encryption protocols such as Secure Sockets Layer (SSL) and its successor Transport Layer Security (TLS) as well as dynamic ports create new challenges for accurate traffic classification, defeating DPI and port-based methods [4]. Classifiers based on classical ML algorithms such as Random Forest, J48 and Bayes Net are considered appropriate to classify encrypted network traffic [4]. However, the traditional ML algorithms require complex feature engineering techniques to reach better accuracy [4]. On the other hand, Deep Learning (DL) can optimize the feature engineering by itself. Moreover, as shown in Section III-D the probability distribution of output layer of Deep Neural Networks (DNN) can be exploited to filter noise and/or unknown data samples.

With this motivation, we propose a novel framework for identifying in-app activities with DNN in encrypted traffic conditions in this paper. Identification of the activities performed by users on mobile apps can be used to profile a mobile user's habits. This is useful for user reconnaissance within networks and in aggregated form, for marketing or intelligence purposes. The proposed method is based on DL, which is a representation learning method [5]. The main contributions of our paper are summarized below.

- There is a wide variety of app categories, from social networking to lifestyle, games, entertainment, health, education, finance etc. It is impractical to train ML algorithms for such a wide range and for each in-app activity in those applications. When deploying an in-app activity classification framework in a real world setting, the framework needs to identify a set of applications and activities within those applications while earmarking previously untrained activities as unknown data traffic. In majority of the existing work in literature, ML algorithms were trained and tested on the same set of applications, which renders them unfit for filtering previously unknown traffic. **The proposed framework in our work is capable of handling network traffic analysis accurately in the presence of noise generated by the unknown traffic.** Further, this framework can provide the percentage of unknown traffic in a given dataset and how the unknown traffic gets misclassified as previously trained in-app activities.
- In attempt to detect in-app activities, there may be instances where an eavesdropper captures the network traffic partially rather than the entire transaction, as user activity may well be already underway [6]. In those cases, most existing work in literature fails to detect an activity if its signature does not fall into the captured window. **The proposed framework can identify fine grained in-app activities even by observing a subset of an activity's traffic.**

Existing work focuses on either coarse grained activities such as browsing, downloading, uploading etc. [7] or generic activities such as posting on Instagram, messaging on WhatsApp etc. [8], [9], [10]. Our research advances the state-of-the-art by identifying fine grained user in-app activities in encrypted network traffic. As such, e.g., when a generic WhatsApp activity 'sending a message' is considered, it can determine whether this message is a long text, short text, image, video, or voice recording. This level of classification is challenging when performed using metadata in an encrypted domain as it requires deep traffic pattern inspection. However, it provides valuable information for an analyst to identify the users, where confidential information is retained. **The proposed framework can identify 92 in-app activities from eight distinct applications.**

- A comprehensive dataset was created by performing a series of actions on apps such as Facebook, Instagram, WhatsApp, Viber, Messenger, Gmail, Skype and YouTube. **To foster new studies and allow reproducing of the results presented, the dataset is shared openly with the research community.** (https://www.dropbox.com/s/9tihcj9wx2sia1t/Dataset.7z?dl=0)

The rest of this paper is organized as follows: In Section II, the related work is reviewed to position the above listed contributions within literature. Section III describes the methodology of the proposed classification system using DL. Sections IV present experimental results and discussions. Finally, Section V concludes the paper.

## II. RELATED WORKS

Conti et al. [8] developed a framework to infer user actions executed on mobile apps based on packet sizes and their order information. Park and Kim [9] target KakaoTalk, a mobile instant messaging service and proposed a framework to infer user activities by passive analysis of network traffic. Saltaformaggio et al. [10] proposed NetScope, a tool to identify user activities generated by mobile apps, based on the statistics originated from the Internet Protocol (IP) headers. The AppScanner [11] framework was implemented for real time identification of Android apps from encrypted network traffic. All these methods employ classical ML algorithms such as k-nearest neighbour and random forest. However, their performance heavily depends on human generated features which is significantly time consuming and limited in generalizability.

DL obviates the need to perform feature selection by a domain expert and it has a higher capacity to learn highly complicated patterns compared to traditional ML methods. Recent work has demonstrated the efficacy of DL methods to perform traffic classification. In [12], a framework named 'Deep Packet' was presented that achieved both traffic characterization, where the network traffic was categorized into major classes (e.g., FTP and P2P), and application identification (e.g., BitTorrent and Skype). Deep Packet framework embedded Stacked Autoencoder (SAE) and One Dimensional Convolutional Neural Networks (1D CNN) to classify network traffic. Experiments were conducted using the ISCX public dataset [13] and the model achieved a recall rate of 0.98 in application identification task and 0.94 in traffic categorization task.

Recurrent Neural Networks (RNN) and CNN were applied to perform application level identification in [14]. Their DL model used packet level features such as ports, payload bytes, TCP window size, inter arrival times of packets and packet direction. This combination of RNN and CNN model attained an accuracy of 96%.

In [5], an end-to-end encrypted traffic classification model with 1D CNN was proposed. Feature extraction, feature selection and classifier were integrated into a single framework in the model, which automatically learnt non-linear relationship between the raw input and expected output. This model was validated with the ISCX public dataset and showed a significant improvement over the C4.5 ML method.

In [15], DataNet was introduced as an application aware framework for smart home networks. It was developed and evaluated using three deep learning based approaches, namely multilayer perceptron, stacked autoencoder, and CNN. Experiments were conducted using the ISCX public dataset with encrypted data samples from 15 applications. The experimental results showed that recall, precision, and f1 score were all greater than 92%.

A Server Name Identification (SNI) classification technique was proposed in [16], using HTTPS features (packet sizes, payload sizes, inter-arrival times, direction). The model consisted of a combination of RNN and CNN and obtained an accuracy of 82.3%.

Aceto et al. proposed two frameworks named MIMETIC [17] and DISTILLER [18]. MIMETIC takes two inputs namely data payload and protocol/time series features. The DL model captures patterns in both input viewpoints to perform traffic classification. Multitask and multimodal DL is adapted to devise the DISTILLER to perform mobile app classification. In [19], FS-Net an end-to-end model was proposed that learns features from the raw flow sequences and makes classification to identify flows. FS-Net achieves a True Positive Rate of 99.14% in identifying 18 applications.

In [20], a traffic classification method based on CNN was proposed. This method used the NetFlow and packet based features to identify Quick UDP Internet Connection (QUIC) protocol based services such as Google Hangout chat, Google Hangout voice call, YouTube, File transfer and Google Play music. The experiments demonstrated that this method could detect five kinds of QUIC based services with an accuracy approximately 99%. This work uses NetFlow based features that leads to an increase in the runtime of processing and classification. Using all packets in the traffic flows is another disadvantage in this work, as this causes obstacles when the number of packets in flows is large.

In [21], a framework called 'ActiveTracker' was proposed to recognize app trajectory over the encrypted Internet traffic streams. Experiments were conducted based on real world encrypted mobile traffic as well as synthetic traffic. The proposed DNN based classification model which consisted of an app filter and an activity classifier achieved up to 79.65% in recognizing app trajectory from a long traffic stream.

Table I presents a summary of the existing body of research in this domain as discussed above. The variety of work in literature primarily focused on classifying previously trained traffic while none has targeted providing network traffic analysis accurately in the presence of noise generated by unknown traffic, albeit this would be a typical situation encountered in a real world scenario. Thus, our work is aimed at advancing the state-of-the-art by identifying previously trained fine grained in-app activities accurately as well as detecting and classifying previously unknown in-app activities as unknown data. As such, the proposed model performs classification based on a few frames of the encrypted traffic flow rather than considering the entire flow of a transaction. The experimental results of this work are compared with those of the state-of-the-art methods [12], [16] to validate the performance of our proposed model.

TABLE I
POSITIONING OF THE PROPOSED WORK WITH RESPECT TO THE EXISTING BODY OF RESEARCH IN LITERATURE

| Research article | Encrypted traffic | Unknown data detection | Classification type | Classification based on subset of traffic | Results compared with the state-of-the-art methods | Dataset type | Accuracy |
|---|---|---|---|---|---|---|---|
| [12] | ✓ | ✗ | Traffic characterization and application | ✗ | ✓ But not DL methods | ISCX | Traffic characterization 94%, app identification 98% |
| [14] | ✓ | ✗ | Service (HTTP, DNS, Telnet, etc.) | ✓ First 20 packets | ✗ | RedIRIS | 96% |
| [5] | ✓ | ✗ | Traffic characterization | ✓ First 784 bytes of a flow session | ✓ But not DL methods | ISCX | 86% |
| [15] | ✓ | ✗ | Application | ✗ | ✗ | ISCX | 93% - 98% precision. accuracy not given |
| [16] | ✓ | ✗ | SNI | ✓ Sequence length 25 | ✗ | Private | 82.3% |
| [17] | ✓ | ✗ | Application | ✗ | ✓ | 3 Private datasets | 79.98%, 89.49%, 89.14% |
| [18] | ✓ | ✗ | Traffic type and application | ✗ | ✓ | ISCX | 93.75% |
| [19] | ✓ | ✗ | Application | ✗ | ✓ | MaMPF | 99.14% True positive rate |
| [20] | ✓ | ✗ | QUIC based services | ✗ | ✗ | Private | 99% |
| [21] | ✓ | ✗ | App trajectory | ✗ | ✓ But not DL methods | Private | 79.65% |
| This paper | ✓ | ✓ | Application and Activity | ✓ | ✓ | Openly shared | 90% to 95% |

### III. PROPOSED METHODOLOGY

In this work, a framework is proposed that comprises deep learning methods, namely DNN for in-app activity identification. This framework consists of four phases, namely the data collection, pre-process, training, and test phases. The pre-process phase transforms the raw traffic data to the required input format for training. At the training phase, a neural network is trained using a labelled training dataset and at the test phase, the pre-trained model's performance is evaluated based on its classification of previously trained and untrained apps' traffic flows.

*A. Data Collection*

A comprehensive dataset was collected through recording a series of user actions performed via eight mobile apps, whose features are presented in Table II. The number of samples for each app is obtained when the traffic flows are segmented into 0.5sec window size.

TABLE II
APP FEATURES RECORDED DURING DATA COLLECTIONS

| App | Category | Number of activities | Number of samples |
|---|---|---|---|
| Facebook | Social networking | 22 | 19,944 |
| Instagram | Photo and video | 20 | 6,818 |
| YouTube | Photo and video | 9 | 14,501 |
| WhatsApp | Social networking | 9 | 436 |
| Viber | Social networking | 9 | 690 |
| Gmail | Productivity | 5 | 1,036 |
| Skype | Social networking | 8 | 15,703 |
| Messenger | Social networking | 10 | 6,061 |
| Total: 8 apps | | 92 | 65,189 |

Data collection for each activity was repeated four times to generate sufficient amount of traffic flows for inference. The number of samples obtained for Facebook, Skype and YouTube is greater. This is because Facebook has lots of activities compared to other apps; YouTube mainly has video related activities such as video watching, uploading, and downloading; Skype has activities such as taking audio and video calls, sending video clips. These activities generate lots of traffic. The captured network traffic was saved as .pcap files. To obtain the ground truth, network traffic generated after executing each activity was collected separately and the network trace was labelled with the name of the activity performed. To minimize traffic noise, only the traffic generated by a target app was captured while no other apps were allowed to run in the background.

In this research, the network was monitored passively, and the network traffic was captured without connecting to the WLAN to which the target user's smartphone was connected. To this end, Airmon-ng and Airodump-ng sniffing tools from the Aircrack-ng [22] suite was used to sniff the network traffic transmitting within a wireless network. Network adapters were set only to capture packets that were sent to them. Therefore, the network adapter was set to the monitoring mode to capture all traffic. Encrypted traffic was sniffed on the same WLAN channel as the access point. Fig. 1 shows the experimental testbed used to sniff traffic. The smartphone was provided access to the Internet over a wireless connection via a router. To avoid other sources of interference, the smartphone was connected to the WLAN exclusively.

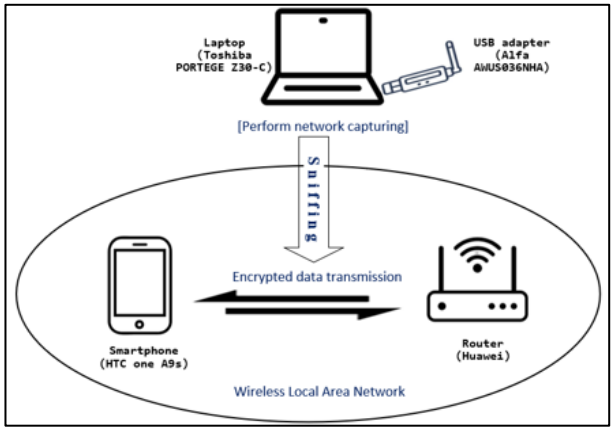

Fig. 1. Test setup for traffic data collection

### B. Data Pre-processing

Before training the model, it is required to prepare the network traffic data so that it can be fed suitably to the neural network. For this, a pre-processing phase was performed on the dataset prepared, with the following steps.

a) Data frame filtering

There are three IEEE 802.11 frame types, namely data, control, and management. Among all, only the data frames are used to carry data during communication, while presence of the other two may hinder the analysis process. Therefore, at this stage management and control frames are eliminated from analysis and remaining data frames are processed further.

b) Obtaining error free frames

Wireless traffic typically suffers from high rates of retransmissions due to packet losses [23]. Packet retransmission may change the traffic pattern of an application. Studies that use statistical features of an entire flow are resistant to a few of the unrelated packets, while methods that use only a few packets to perform classification may be affected more [24]. Since our work is based on using only a part of the activity related traffic for classification, the retransmission frames were filtered out and only error free frames were processed further.

c) Data normalization

Data normalization is a step that is crucial to DL performance. Feature scaling aka standardization is performed to normalize the data. The dataset contains feature values that are different in scale. The variables with larger ranges will dominate over those with small ranges, leading to biased results. Therefore, to equalize the importance of all features, standardization is applied on the feature values to bring them into the same scale. This also allows DL methods to converge faster. Standard scaler [25] was used to standardize a feature by subtracting the mean and then scaling to unit variance. Below formula is used to perform the standardization:

$$X' = (X - \mu) / \sigma \qquad (1)$$

where, X' is the normalized value of the feature, X is the original value, $\mu$ and $\sigma$ are the mean and standard deviation of the feature values, respectively.

d) Traffic Segmentation

During in-app activity detection, it is not possible to ensure that the entire transaction of a user activity may be observed. This is because there can be instances where the eavesdropper may start capturing the traffic while the target user is in the middle of performing an activity. In these situations, the eavesdropper will observe only a part of the transaction instead of its entirety. Therefore, instead of considering the traffic generated by the entire user transaction, a short window was used to divide the traffic into shorter segments. In this work, different time window sizes were set, and the classification accuracies were checked against those. The obtained results are presented in Section IV-A. This provides the opportunity for an eavesdropper to detect in-app activities even by observing only a part of an activity's traffic.

### C. Neural Network Architecture

In this work, a DNN has been used for performing necessary classifications. As shown in Fig. 2, the first layer of the DNN is an input layer which contains the set of features. The two characteristics of traffic considered in this research were the frame length and frame inter arrival time measured in bytes and seconds, respectively. For each of these 12 different statistics (minimum, maximum, standard deviation, first quartile, second quartile, third quartile, mean, median absolute deviation, variance, skew, kurtosis, sum) were calculated separately in the uplink and downlink channels. Hence, in total 48 features (12 statistics x 2 directions x 2 characteristics) were considered as input to the DNN (see the input layer Fig. 2). Python Pandas library [26] was used to compute these features. The final output layer is a layer of nodes that produces the output variables in a DNN setup. In this work, the output layer consisted of 92 nodes which is equal to the number of in-app activities considered in this work (see the output layer Fig. 2).

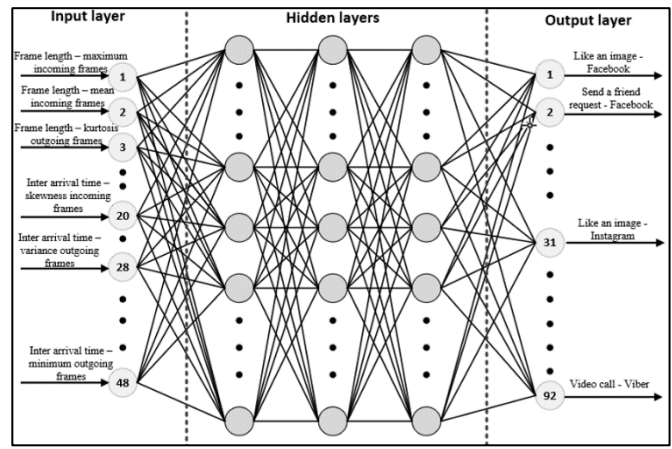

Fig. 2. Neural network architecture.

Construction of a neural network involves several critical decisions such as the number of hidden layers, number of nodes in each hidden layer, choice of activation function etc. The performance of a model can vary considerably according to the selected set of hyper parameters. There is no rule for deciding the hyperparameters of a DNN model. Therefore, a series of tests were performed with different hyperparameter values. It was observed that when the complexity of the model is increased with increasing number of hidden layers and node values, the time it takes to train the model also increases. However, when the structure is simple it suffers from reduced training and validation accuracies. After conducting several experiments and analyses, we present the results obtained from the following DNN architecture in Table III. This architecture can be optimized to increase the accuracy.

TABLE III
THE RESULTING MODEL

| Parameter | Values |
| --- | --- |
| Number of hidden layers | 4 |
| Number of nodes in each layer | [1024, 512, 256, 128] |
| Activation functions | Hidden layers – Tanh
Output layer - Softmax |
| Loss function | Categorical cross entropy |
| Optimizer | Adam |
| Batch size | 2048 |
| Number of epochs | 100 |

The proposed DNN model architecture consists of four hidden layers made up of 1024, 512, 256 and 128 neurons. All layers employ Tanh as the activation function, except for the final softmax classifier layer. To prevent overfitting, the dropout technique was employed with a 0.3 dropout rate. During the training phase, the dropout technique randomly sets a series of the neurons to zero. Therefore, at each iteration there is a random set of active neurons [12].

### D. Classification Technique to Identify Untrained In-app Activities

Identifying untrained in-app activities as unknown traffic is one of the key contributions of this work. For this purpose, the trained model given in Table III is used and the probability distribution of the output layer has been exploited. The technique used to detect noise generated by previously untrained in-app activities is shown in Fig. 3.

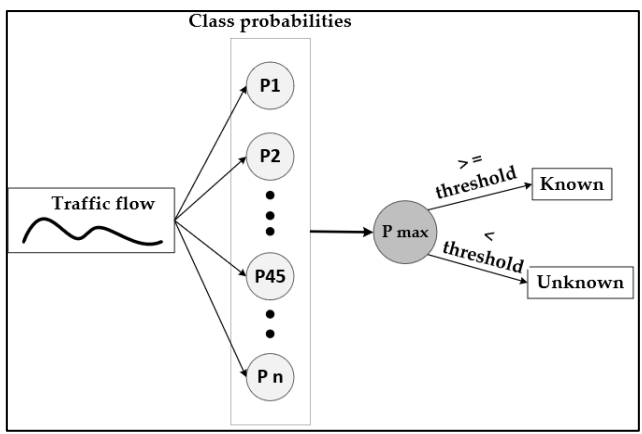

Fig. 3. Noise detection framework.

When input traffic is presented, the model's output layer can predict the probability of the data instance belonging to each class. The number of nodes in the output layer is equal to the number of classes in the pre-trained model. Let us denote this as N_out. Hence, for each input data, the DNN architecture outputs N_out number of values. The values are between 0 and 1, and since these are probabilities, they add up to 1. Getting a value 1 and 0 represents the likelihood of the observation belonging to a class is 100% and 0%, respectively. Typically, for any given input, one node is expected to have a value higher than the rest of the other output nodes. In Fig. 3 the node with the highest probability is denoted as P max. The decision for converting a predicted probability into a class label is governed by a parameter referred to as the threshold. If the P max is less than the threshold value, then the input traffic flow is considered as an unknown instance. The reason for this is that the model is not highly confident. By setting a threshold on the positive class, it determines whether the input data belongs to one of those trained output nodes, which translates into one of the pre-trained in-app activities or not.

To test the impact of threshold value on model's performance, a range of threshold values were selected, and tests were performed. The threshold value that contributes to achieving the highest classification accuracy was selected as reported in Section IV-C.

Use of threshold at the output layer to exploit the phenomenon is novel. Because typically in classification tasks, the node with the highest probability will be chosen and the corresponding class label gets assigned. In this research, we use a two stage approach to check if the node with the highest probability satisfies a pre-set threshold value. Based on this, the known and unknown instances get classified. However, setting this threshold is challenging, as setting it too high increases false negatives whereas setting it too low increases the false positives. Refer to Section IV-C for further details.

Parallel to this work, we also tested the performance of this approach using CNN architecture in our previous work [29].

## IV. EXPERIMENTAL RESULTS AND DISCUSSIONS

In this section, performance evaluation of the proposed framework is presented. The performance evaluation is twofold:

- When previously trained in-app activities are input to the model, it should be able to detect them accurately.
- When previously untrained in-app activities are input to the model, it should be able to identify them as unknown data.

To develop the proposed DNN, the Keras library [27] was used with TensorFlow [28] as its backend. The training dataset consisted of network traffic from in-app activities of the Facebook, Instagram, YouTube, Viber, WhatsApp, Gmail, Skype and Messenger apps. The training dataset was split into two subsets: the first one which included 80% of the total training samples was used for training while the second contained the remaining 20% of the samples was used for validation. The validation performed during the training phase aimed to determine how well the model could identify previously trained in-app activities. In the testing phase, the model's capability to identify previously untrained in-app activities as unknown data was checked.

To evaluate the performance of the trained model, the accuracy metric was computed by:

$$Accuracy\ of\ known\ data = \frac{True\ Positives + True\ Negatives}{Total\ no.of\ known\ instances} \quad (2)$$

$$Accuracy\ of\ unknown\ data = \frac{True\ Negatives}{Total\ no.\ of\ unknown\ instances} \quad (3)$$

### A. In-app Activity Detection

The DNN model architecture proposed in Table III was used to perform in-app activity identification. Instead of considering the entire traffic flow generated by a user transaction, traffic was divided into shorter segments. In this work, four different time window sizes were considered to divide the flows. They are 0.5sec, 0.2sec, 0.05sec and 0.02sec. Classification accuracies obtained for the 92 in-app activities were checked against these window sizes and are reported in Table IV.

TABLE IV
IN-APP ACTIVITY CLASSIFICATION PERFORMANCE

| Window size | Total samples | Training accuracy | Validation accuracy |
|---|---|---|---|
| 0.5 sec | 65,189 | 97% | 93% |
| 0.2 sec | 231,824 | 97% | 95% |
| 0.05 sec | 1,876,624 | 94% | 91% |
| 0.02 sec | 7,067,363 | 93% | 90% |

From the experimental results, it was observed that the validation accuracy is the highest when the window size is 0.2sec. This is because when the flows are split into 0.2sec sizes, we were able to obtain plenty of samples compared to the 0.5sec split. Training the model with a greater number of samples led to an increase in accuracy. However, when the traffic flows were further split into 0.05sec and 0.02sec windows, even though the number of samples increased, we can see the accuracies have decreased. This is because when the splitting window size gets smaller, the number of frames in those window segments also reduces. When frames are considered individually [7] or when the number of frames is small, they contain very little information leading to a direct negative effect on the accuracies. Furthermore, when the split size gets smaller, time it takes to train and test the models increases. This is because when the window size is smaller, the traffic flows are divided into larger number of segments, which take more time to process and train the produced traffic flows. However, for all the above mentioned window sizes, the model was able to obtain an accuracy of 90% or above. This shows the possibility of the proposed model's ability to identify in-app activities even by observing only a small subset of an activity's traffic.

### B. Unknown In-app Activities Detection

In this work, a total of eight apps are considered. From the collected dataset, two separate sub datasets: training dataset and test dataset were created. At each instance, one of the apps was selected and separated to be used in the test dataset and the remaining seven apps were used in the training dataset. Training dataset was used to train the model. To test how well the model was capable of detecting "unknown" traffic, the test dataset with the previously untrained in-app activities was input to the model. Values obtained at each test instance are seen in Table V.

TABLE V
NOISE DETECTION RATES OF UNKNOWN IN-APP ACTIVITIES

| Test no. | The apps used to train the DNN model. | The app not used in model training. | Noise detection |
|---|---|---|---|
| T1 | I, Y, G, M, S, W, V | F | 88% |
| T2 | F, I, G, M, S, W, V | Y | 62% |
| T3 | F, Y, G, M, S, W, V | I | 75% |
| T4 | F, I, Y, M, S, W, V | G | 91% |
| T5 | F, I, Y, G, S, W, V | M | 75% |
| T6 | F, I, Y, G, M, W, V | S | 63% |
| T7 | F, I, Y, G, M, S, W | V | 91% |
| T8 | F, I, Y, G, M, S, V | W | 85% |

In Table V, Facebook, YouTube, Viber, Gmail, Skype, Messenger, Instagram, and WhatsApp apps are denoted by F, Y, V, G, S, M, I and W letters, respectively. Considering all the tests reported in the table, the model achieved an average of 79% in detecting noise. When Gmail and Viber were input to the model as test traffic, most of its traffic got correctly classified as unknown resulting a high noise detection rate. However, most of the YouTube and Skype test traffic got misclassified as one of the already trained classes resulting in a low noise detection rate. In test T2, 62% of the YouTube traffic has been detected correctly as unknown traffic/noise. The remaining 38% of the traffic was incorrectly classified as activities that contain in the training dataset.

To have a better insight into the nature of the misclassified traffic, the apps to which test (unknown) traffic got classified were further analysed. The results are presented in Table VI in percentages. For example, 12% of the data from F is classified as known data (see T1 in Table V). The distribution of this 12% of the misclassified data from F is shown in the first row in Table VI. Majority (34% and 33%) of this data are assigned to S (Skype) and Y (YouTube) respectively. The apps that contribute to the misclassification with the highest percentage value are shaded in dark blue and those that have the low percentages are in light blue.

TABLE VI
CONFUSION MATRIX REPRESENTING MISCLASSIFICATIONS (% VALUES)

| | | F | I | Y | G | M | S | W | V |
|---|---|---|---|---|---|---|---|---|---|
| Test apps | F | | 15 | 33 | 2 | 13 | 34 | 1 | 2 |
| | I | 34 | | 25 | 2 | 10 | 27 | 1 | 1 |
| | Y | 41 | 11 | | 2 | 12 | 31 | 1 | 2 |
| | G | 31 | 10 | 22 | | 10 | 25 | 1 | 1 |
| | M | 35 | 11 | 24 | 2 | | 26 | 1 | 1 |
| | S | 41 | 13 | 29 | 2 | 12 | | 1 | 2 |
| | W | 31 | 11 | 22 | 2 | 9 | 24 | | 1 |
| | V | 31 | 10 | 22 | 2 | 9 | 25 | 1 | |

Looking at Table VI, the majority of the misclassified data was assigned to Facebook. More than 33% of the misclassified traffic from YouTube, Skype, Messenger, and Instagram test apps belongs to Facebook. The reason for this is shown in Fig. 4 (a), where the corelation graphs of Facebook vs those four apps are presented. In Fig. 4(a) the points are closely packed. This means the strength of the correlation is high among these apps with Facebook, which has caused them to get misclassified as Facebook traffic. Facebook Inc.'s apps (i.e., Facebook, Instagram, and Messenger) use FB-Zero protocol. Most of misclassifications occur within the same protocol group, that is wrongly assigning to a flow a label of an app using the same protocol.

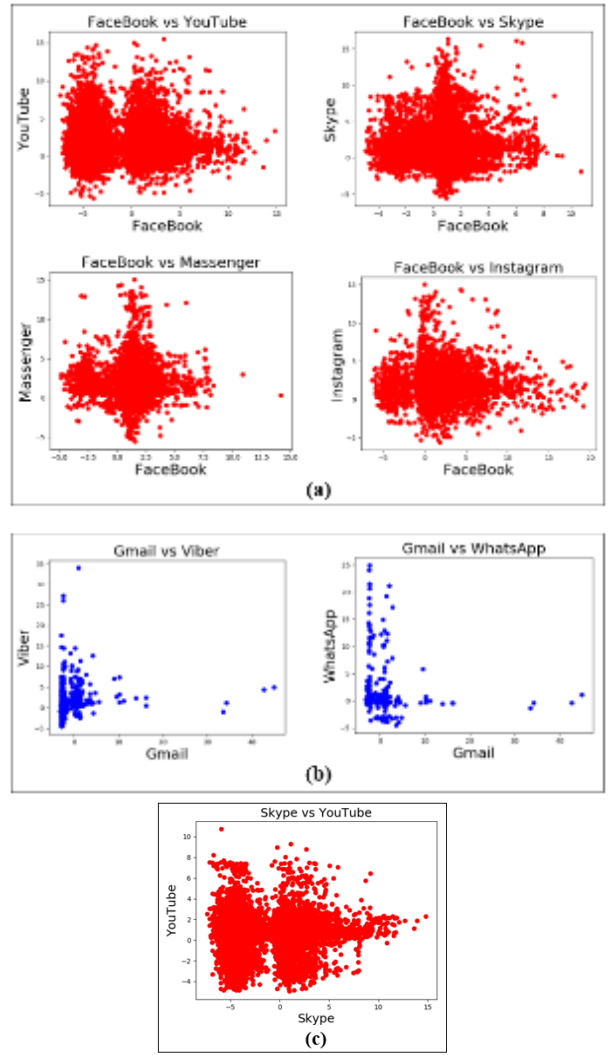

Fig. 4. Correlation graphs of applications (a) correlation of Facebook with YouTube, Skype, Messenger and Instagram. (b) correlation of Gmail with Viber and WhatsApp. (c) correlation of Skype vs YouTube.

Apps that are highly correlated with each other have similar in-app activities with similar behaviour. For example, when Instagram was input to the model as the test data, majority of (34%) of the misclassified traffic was labelled as Facebook. This is because, both these social networking apps provide similar functionalities such as posting on wall, liking posts, commenting on posts, adding stories etc. Therefore, it is likely that having similar in-app activities to cause the Instagram samples to be misclassified as part of the Facebook traffic.

To get a better insight into the misclassification, the in-app activities that cause such misclassification among these apps are considered. When YouTube is the test app, the highest number of misclassified instances comes from Facebook live in-app activity (21%), as YouTube has a similar activity, which is uploading a video. 86% of the YouTube traffic is from this activity. For Skype, Messenger, and Instagram, the highest number of misclassified instances 25%, 21%, 21% respectively come from YouTube video uploading in-app activity. Video support is available in all the three test apps, and thus having such similar in-app activity has caused the misclassification.

When Gmail is considered as the test app, only 9% of the unknown traffic was misclassified as previously trained class labels. 1% of the misclassified traffic was labelled as traffic belonging to WhatsApp, and another 1% was labelled as traffic belonging to Viber. This means that the lowest misclassified Gmail unknown traffic came from WhatsApp and Viber. On the other hand, when WhatsApp and Viber are considered as test apps, 2% of the misclassified traffic was labelled as Gmail. This is because of the inherently distinctive dissimilarities in the activities between Gmail-Viber and Gmail-WhatsApp. This means that the correlation of Viber and WhatsApp with Gmail is low. This has been demonstrated by the correlation graphs of Fig. 4 (b). It can be observed that the points are distributed loosely which means the considered apps are only slightly correlated to each other.

When Skype is provided as the unknown test dataset to the model, 37% of Skype traffic was misclassified as trained class labels. 29% of the misclassified traffic was labelled as YouTube. After Facebook, this is the second highest misclassified class label of Skype. On the other hand, when YouTube is considered, 38% of the unknown YouTube test traffic was misclassified as trained class labels. 31% of the misclassified traffic was labelled as Skype. This is the second highest misclassified class label of YouTube. To understand the cause of the misclassification between Skype and YouTube, we obtained the corelation graphs of the two apps. As shown in Fig. 4 (c), a high correlation between Skype and YouTube can be noted, which has caused this misclassification.

From the experiments reported above to detect unknown traffic, it is evident that the proposed model can effectively identify previously untrained in-app activity traffic as unknown data with an average accuracy of 79%. Misclassification is caused by apps that have high correlation among each other due to their similar network behaviours and having similar in-app activities. Apps from the same provider and sharing the same underlying protocols also cause traffic misclassification.

### C. Threshold Selection

To obtain a calibrated and robust indication of the model's performance, the confidence distribution of all the known and unknown data instances were examined. The probability output by the model's output layer is used to obtain the confidence distribution of instances. For any given input instance, one node in the model would result in having a probability value higher than the rest of the output nodes. This is refereed as the confidence value. The number of known and unknown instances that falls under each confidence value is obtained to plot the histogram depicted in Fig. 5. This demonstrates that the majority of the input data associated with the known data (orange coloured bars) is skewed towards the higher probability, the majority of unknown data is distributed uniformly in the lower probabilities. This demonstrates that setting a threshold at probability closer to 1 will be able to remove majority of unknown data. Following tests with several threshold values, it was observed that when the threshold value was set to 0.97, the model reached the most optimum operating point characterized by the set of highest training accuracy, validation accuracy and unknown data detection rate. This is detectable in the area indicated by the red bounding box in the Fig. 5, where the selected threshold value resulted in the highest number of known and unknown data samples. This means most of the positive instances showed a high confidence level during the classification phase. Higher the confidence level of the instances, greater the accuracy of the results obtained from the model.

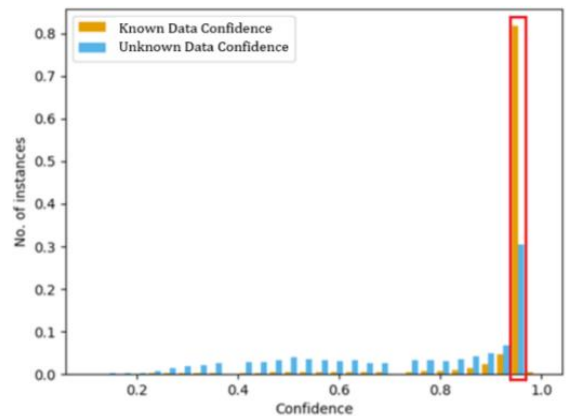

Fig. 5. Confidence distribution of training and testing data.

### D. Performance Comparison with the State-of-the-art

For the performance comparison of our framework with the state-of-the-art, two best matching models have been considered, namely Bayat et al. model and Deep Packet model. For a fair comparison, the same dataset generated in our research was also used for traffic classification employed in these selected DL models.

The model proposed by Bayat et al. [16] consists of a combination of RNN and CNN. The first two layers are 1D convolutional layers with 200 and 400 filters respectively, which employ the ReLU activation function. Each 1D CNN layer is followed by a batch normalization to speed up the learning phase. Next, Gated Recurrent Units (GRU) with 200 hidden units is included to process sequences of variable length. GRU layer is followed by two fully connected layers with 200 and 63 (matching the number of classes in our work) hidden units, respectively. Sigmoid and Softmax activation functions are used for the two fully connected layers, respectively. Dropout between the fully connected layers is used to reduce over fitting. The model was trained for 100 epochs and used a batch size of 64, Adam optimizer and categorical cross entropy loss. This model is used in this research to perform in-app activity classification. When this model was tested with the dataset generated in this study, an accuracy of 81% was obtained.

The Deep Packet model presented by Lotfollahi et al. [12] comprises two deep learning methods, namely CNN and SAE. The SAE architecture consists of five dense layers with 400, 300, 200, 100 and 50 neurons respectively. All five dense layers employ the ReLU activation function. After each dense layer, the dropout technique is employed with 0.05 dropout rate to avoid over fitting. In the final layer, a Softmax classifier with 63 neurons (matching the number of classes in our work) is applied for the classification task. The 1D CNN model of [12] consists of two convolutional layers with 200 filters, followed by a pooling layer. Then, the Two Dimensional (2D) tensor is squashed into a 1D vector and fed into a network of three dense layers which also employ the dropout technique. All these layers employ the ReLU activation function. Finally, a Softmax classifier is applied to perform the classification. For this CNN model, the number of neurons of the three dense layers are not provided. Therefore, the number of neurons of the first three dense layers (1024, 512, 256) of our proposed model have also been utilized for the performance comparison tests.

In both models considered in [12], batch normalization was used to speed up the learning phase. Categorical cross entropy and Adam were used as the loss function and optimizer, respectively. SAE model was trained for 200 epochs and CNN was trained for 300 epochs. The batch size was not available in this work, thus the batch size obtained for the model of our framework was also used here, which was 2048. These models are used in this research to perform in-app activity classification. When the SAE and CNN models were tested with the dataset generated in this study, accuracies of 34% and 32% were obtained, respectively.

TABLE VII
PERFORMANCE COMPARISON WITH THE STATE-OF-THE-ART

| Paper | Architecture | Accuracy with original dataset | Accuracy with our dataset |
|---|---|---|---|
| [16] | 1D CNN and RNN | 82.3 % | 81% |
| [12] | 1D CNN | 98% | 32% |
|  | SAE | 95% | 34% |
| This paper | DNN | 93% | 93% |

Table VII shows the accuracy values obtained from the models introduced in [16] and [12] when they are tested using their original datasets reported vs when tested using the dataset generated in our study. The datasets used in the referenced works were not publicly available for our use to test our model. The results presented show that there is a significant decrease in accuracies when the two selected state-of-the-art models are used to perform in-app activity identification using the dataset generated in this research. As these models consider the same data to perform training and testing, they suffer when unknown data is presented.

In this work the dataset is collected, and the model is tested in an ideal setting. However, when this model is deployed in a real world traffic scenario, such as in a coffee shop or airport, we except that there will be differences in model accuracies, but without any major variations in detecting and eliminating unknown traffic effectively.

V. CONCLUSIONS

This work contributes to the literature in encrypted traffic classification by applying a DNN architecture for fine grained in-app activity detection. Experiments were executed to find out the hyperparameters that maximize the performance of the DL model. Results showed that the proposed model effectively identified previously trained in-app activities with an accuracy of 90% or above while filtering out previously untrained in-app activity traffic as unknown data with an average accuracy of 79%. The proposed windowing approach enables the model to perform classification based on a few frames of the traffic flow instead of the entire flow of a transaction. This is significantly a harder task, since there is more information in the entire traffic flow compared to a part of the flow. This feature allows our model to be more applicable in real world situations. Because when detecting in-app activities, there may be instances where an eavesdropper starts capturing the traffic while the target user's activity is already underway, which leads to observing only a part of the activity related traffic. Further, classifications were performed only considering side channel data statistics, without decryption or using any header information. The novel approach of using threshold on the confidence values exploits the output layer of the trained DNN model to identify in-app activities while removing unknown in-app activity traffic.

As future work, we plan to convert the encrypted traffic flow sequences into multi channel images, and then to apply DL models on the generated images to perform traffic classification.

REFERENCES

[1] Flurry Analytics, comScore, Pandora, Facebook, NetMarketShare, https://www.marketingcharts.com/industries/media-and-entertainment58693/attachment/ flurry-share-time-spent-mobile-devices-sept 2015, 2021.
[2] Al-Mejibli, I.S. and N.R. Alharbe, "Analyzing and evaluating the security standards in wireless network: a review study". Iraqi Journal for Com-puters and Informatics, 46(1), pp.32-39, 2020.
[3] V. F. Taylor, R. Spolaor, M. Conti, and I. Martinovic, "Robust smartphone app identification via encrypted network traffic analysis," IEEE Transac-tions on Information Forensics and Security, 13(1), pp.63-78, 2017.
[4] G. Aceto, D. Ciuonzo, A. Montieri and A. Pescapé, "Mobile Encrypted Traffic Classification Using Deep Learning," Network Traffic Measure-ment and Analysis Conference (TMA), Vienna, pp. 1-8, doi: 10.23919/TMA.2018.8506558, 2018.
[5] W. Wang, M. Zhu, J. Wang, X. Zeng and Z. Yang, "End-to-end encrypted traffic classification with one-dimensional convolution neural net-works," IEEE International Conference on Intelligence and Security In-formatics (ISI), Beijing, pp. 43-48, doi: 10.1109/ISI.2017.8004872, 2017
[6] M.H. Pathmaperuma, Y. Rahulamathavan, S. Dogan, and A.M. Kondoz, "In-App Activity Recognition from Wi-Fi Encrypted Traffic," in: Arai K., Kapoor S., Bhatia R. (eds) Intelligent Computing. SAI 2020. Advances in Intelligent Systems and Computing, vol 1228. Springer, Cham. https://doi.org/10.1007/978-3-030-52249-0_46, 2020.
[7] Zhang, F., He, W., Liu, X., And Bridges, P. G. "Inferring users' online ac-tivities through traffic analysis". In Proc. ACM Conference on Wireless Network Security, 2011.
[8] M. Conti, L. V. Mancini, R. Spolaor, and N. V. Verde, "Analyzing Android encrypted network traffic to identify user actions," IEEE on Information Forensics and Security, vol. 11, no. 1, pp.114–125, January 2016.
[9] K. Park and H. Kim, "Encryption is not enough: Inferring user activities on KakaoTalk with traffic analysis," in Proceedings of the 16th Interna-tional Workshop on Information Security Applications, ser. WISA '15, H.-W. Kim and D. Choi, Eds. Berlin, Heidelberg: Springer, pp. 254–265, 2015.
[10] Saltaformaggio, B., Choi, H., Johnson, K., Kwon, Y., Zhang, Q., Zhang, X., Xu, D., Qian, J.: "Eavesdropping on fine-grained user activities within smartphone apps over encrypted network traffic". In: WOOT'16 Pro-ceedings of the 10th USENIX Conference on Offensive Technologies, pp 69-78. USENIX Association Berkeley, CA, USA, 2016.
[11] V. F. Taylor, R. Spolaor, M. Conti, and I. Martinovic, "AppScanner: Auto-matic Fingerprinting of Smartphone Apps from Encrypted Network Traffic," in IEEE European Symposium on Security and Privacy (Euro S&P), pp. 439–454, March 2016.
[12] M. Lotfollahi, R. S. H. Zade, M. J. Siavoshani, and M. Saberian, "Deep packet: A novel approach for encrypted traffic classification using deep learning". Soft Computing, 24(3), pp.1999-2012, 2020.
[13] G. Draper-Gil, A. H. Lashkari, M. S. I. Mamun, and A. Ghorbani, ''Char-acterization of encrypted and VPN traffic using time-related features'', in Proc. ICISSP, pp. 407–414, 2016.
[14] M. Lopez-Martin, B. Carro, A. Sanchez-Esguevillas, and J. Lloret, "Net-work traffic classifier with convolutional and recurrent neural networks for internet of things", IEEE Access, 5:18042–18050, 2017.
[15] P. Wang, F. Ye, X. Chen, and Y. Qian, "DataNet: Deep learning based en-crypted network traffic classification in SDN home gateway", IEEE Ac-cess, vol. 6, pp. 55380–55391, 2018.
[16] N. Bayat, W. Jackson, and D. Liu, "Deep Learning for Network Traffic Classification", arXiv preprint arXiv:2106.12693, 2021.

[17] G. Aceto, D. Ciuonzo, A. Montieri, and A. Pescapé, "MIMETIC: Mobile encrypted traffic classification using multimodal deep learning". Computer Networks, 165, p.106944, 2019.

[18] G. Aceto, D. Ciuonzo, A. Montieri, and A. Pescapé, ''DISTILLER: Encrypted traffic classification via multimodal multitask deep learning,'', Journal of Network and Computer Applications, 183, p.102985, 2021.

[19] C. Liu, L. He, G. Xiong, Z. Cao, Z. Li, "FS-Net: A Flow Sequence Network for Encrypted Traffic Classification", IEEE Conference on Computer Communications (INFOCOM), 1171–1179, 2019.

[20] V. Tong, H.-A. Tran, S. Souihi, and A. Mellouk, "A novel QUIC traffic classifier based on convolutional neural networks", In 2018 IEEE Global Communications Conference (GLOBECOM) (pp. 1-6), 2018.

[21] Li, D., Li, W., Wang, X., Nguyen, C.T. and Lu, S., "App trajectory recogni-tion over encrypted internet traffic based on deep neural network," Computer Networks, 179, p.107372, 2020.

[22] Aircrack-ng, https://www.aircrack-ng.org/, 2021.

[23] M. Rodrig, C. Reis, R. Mahajan, D. Wetherall, and J. Zahorjan, "Measure-ment-based characterization of 802.11 in a hotspot setting", in: Proceed-ings of the ACM SIGCOMM Workshop on Experimental Approaches to Wireless Network Design and Analysis, pp. 5 -10, 2005.

[24] S. Rezaei and X. Liu, "Deep Learning for Encrypted Traffic Classification: An Overview," in IEEE Communications Magazine, vol. 57, no. 5, pp. 76-81, doi: 10.1109/MCOM.2019.1800819, May 2019.

[25] Sklearn.preprocessing. StandardScaler, http://scikit-learn.org/stable/modules/generated sklearn.preprocessing. StandardScaler.html., 2021

[26] W. McKinney, "Data Structures for Statistical Computing in Python", Proceedings of the 9th Python in Science Conference, pp. 51-56, 2010.

[27] F. Chollet, et al., Keras, https://github.com/fchollet/keras, 2020.

[28] M. Abadi, et al., "TensorFlow: Large- scale machine learning on hetero-geneous systems". URL http://tensorflow.org/, software available from tensorow.org, 2015.

[29] MH Pathmaperuma, Y Rahulamathavan, S Dogan, A Kondoz. CNN for User Activity Detection Using Encrypted In-App Mobile Data. Future Internet. 2022; 14(2):67. https://doi.org/10.3390/fi14020067